\documentclass[aps,twocolumn,prd,superscriptaddress,nofootinbib,floatfix]{revtex4-2}

\usepackage{mathtools}
\usepackage{amsfonts}
\usepackage{amssymb}
\usepackage{mathrsfs}
\usepackage{bbm}
\usepackage{slashed}
\usepackage{amsmath}
\usepackage{bm}
\usepackage{bbold}

\usepackage{graphicx}
\usepackage{color}
\usepackage{colortbl}
\usepackage{array}

\usepackage{float}
\usepackage{placeins}
\usepackage{booktabs}
\usepackage[caption=false]{subfig}
\usepackage{makecell}
\usepackage{tabstackengine}

\usepackage{xspace}
\usepackage{hyperref}
\usepackage[nameinlink]{cleveref}
\usepackage{bookmark}
\usepackage{siunitx}

\usepackage{xifthen}
\usepackage{xcolor}
\hypersetup{
	colorlinks,
	linkcolor={red!75!black},
	citecolor={blue!75!black},
	urlcolor={blue!75!black}
}

\usepackage[utf8]{inputenc}
\allowdisplaybreaks[4]

\newcommand{\Tr}{{\text{Tr}}}

\newcommand{\feyn}[1]{
	\setbox0=\hbox{\ensuremath{#1}}
	\hbox to\wd0{\hbox to0pt{\hbox to\wd0{\hss/\hss}\hss}\box0}}

\def\Eq#1{Eq.~\labelcref{#1}}

\def\Fig#1{Fig.~\labelcref{#1}}

\def\sec#1{Sec.~\labelcref{#1}}
\def\app#1{App.~\labelcref{#1}}

\setkeys{Gin}{width=0.48\textwidth}
\graphicspath{{./figures/}}

\newcommand{\gettitle}{}

\newcommand{\getFudanAffiliation}{\affiliation{Key Laboratory of Nuclear Physics and Ion-beam Application (MOE),
and Institute of Modern Physics, Fudan University, Shanghai 200433, P.R. China}}

\newcommand{\getNSFCFudanAffiliation}{\affiliation{Shanghai Research Center for Theoretical Nuclear Physics, NSFC and Fudan University, Shanghai 200438, P.R. China}}

\newcommand{\getDalianAffiliation}{\affiliation{School of Physics, Dalian University of Technology, Dalian, 116024, P.R. China}}

\newcommand{\getGiessenAffiliation}{\affiliation{Institut f\"ur Theoretische Physik, Justus-Liebig-Universit\"at Gie\ss en, 35392 Gie\ss en, Germany}}

\hypersetup{
	colorlinks,
	linkcolor={red!75!black},
	citecolor={blue!75!black},
	urlcolor={blue!75!black},
	pdftitle={\gettitle},
	pdfauthor={Zelle},
	pdfkeywords={effective theory} {analytic continuation}
	{correlations functions} {hadronization}
	{functional renormalisation group}
	{real time} {bound states}
	bookmarksopen=true,
	bookmarksopenlevel=2,
	bookmarksnumbered=true
}

\begin{document}

\title{High-order fluctuations of temperature in hot QCD matter}
	
\author{Jinhui Chen}
\email{chenjinhui@fudan.edu.cn}
\getFudanAffiliation
\getNSFCFudanAffiliation 

\author{Wei-jie Fu}
\email{wjfu@dlut.edu.cn}
\getDalianAffiliation

\author{Shi Yin}
\email{shiyin.dalian@gmail.com}
\getGiessenAffiliation

\author{Chunjian Zhang}
\email{chunjianzhang@fudan.edu.cn}
\getFudanAffiliation
\getNSFCFudanAffiliation

\begin{abstract} 

A new thermodynamic state function is introduced to describe the thermodynamics relevant for the mean transverse momentum fluctuations of charged particles in heavy-ion collisions, which allows us to compute the temperature fluctuations of different orders in hot quantum chromodynamics (QCD) matter for the first time. Consequently, it is found that the temperature fluctuations are suppressed remarkably as the system transitions from the hadron resonance gas (HRG) to the quark-gluon plasma (QGP) with increasing temperature or baryon chemical potential, alongside a negative skewness. This is attributed to the general fact that the heat capacity of QCD matter increases significantly in QGP in comparison to that in HRG. These predictions provide a candidate observable to discover the thermodynamic temperature fluctuations in upcoming heavy-ion collision experiments, which also paves a novel way to study QCD thermodynamics and QCD phase diagram through measurements of the mean transverse momentum fluctuations of charged particles.

\end{abstract}
	
\maketitle

\section{Introduction}
\label{sec:introduction}

The temperatures and densities reached in relativistic heavy-ion collisions are expected to be similar to those thought to have prevailed in the very early universe, prior to the formation of protons and neutrons. The observation and study of QCD matter, the quark-gluon plasma (QGP) under these conditions has been one of the main science goals of high energy nuclear physics facilities \cite{Shuryak:1980tp, Harris:1996zx, STAR:2005gfr, PHENIX:2004vcz, ALICE:2005vhb, Busza:2018rrf, STAR:2023jdd}. Many questions have been addressed and developed in heavy-ion research programs \cite{Busza:2018rrf}. The occurrence of a phase transition from the QGP to a hadron resonance gas (HRG) or the existence of a critical end point (CEP) in the phase diagram of QCD matter is an exciting scientific endeavor~\cite{Stephanov:1998dy, Stephanov:1999zu, Fu:2019hdw, Gao:2020fbl, Gunkel:2021oya}, which may be revealed by measurements of thermodynamic fluctuations \cite{STAR:2010vob, Bzdak:2019pkr, Chen:2024aom}, such as net-baryon or net-proton number fluctuations \cite{STAR:2020tga, STAR:2021fge, STAR:2022vlo,STAR:2022etb, STAR:2025zdq, Fu:2016tey, Fu:2021oaw, Fu:2023lcm, Lu:2025cls} and temperature fluctuations \cite{Gavin:2003cb,Cao:2021zhy}. Event-by-event (EbE) fluctuations in charged particle momentum distributions serve as probes of thermalization and the statistical nature of particle production in such collisions \cite{Heiselberg:2000fk, Jeon:2000wg, Voloshin:1999yf, Asakawa:2000wh,Stodolsky:1995ds}.  

Recent advances in heavy-ion collision experiments now enable the isolation of the thermal fluctuations from confounding effects, such as the initial state geometry fluctuations \cite{Gardim:2011xv,Schenke:2014tga, STAR:2024wgy,ATLAS:2024jvf,Zhang:2025yyd,STAR:2025elk}, flow contributions, and other non-thermal sources, allowing the direct measurements of thermodynamical properties of hot QCD matter \cite{Gardim:2019xjs,Liu:2025fbu}, e.g., the speed of sound \cite{Gardim:2019brr, Gardim:2024zvi, CMS:2024sgx, Mu:2025gtr}. High-order thermodynamical quantities, e.g., the temperature fluctuations, could be used to probe the QCD thermodynamics and phase transitions, since they are more sensitive to the critical fluctuations, i.e. the singular part of the thermodynamic potential during the phase transition in comparison to low-order thermodynamical quantities, such as the temperature itself \cite{NA60:2008dcb, HADES:2019auv, Churchill:2023zkk, STAR:2024bpc, Chen:2026gka}. This is similar to the case of ordinary net-proton number fluctuations. In principle, temperature fluctuations can be extracted from EbE mean transverse momentum fluctuations of final-state charged particles \cite{Gavin:2003cb}, isolated from other effects, e.g., the hydrodynamic effect \cite{Giacalone:2020lbm}. To that end, one has to study the thermodynamical properties of temperature fluctuations from the theoretical side, in particular the smoking-gun signature of temperature fluctuations, in measurements of EbE mean transverse momentum fluctuations that have been extensively done across collision energies and systems in various heavy-ion facilities \cite{NA49:1999inh, CERES:2003sap, PHENIX:2003ccl, STAR:2005vxr, NA49:2008fag, ALICE:2023tej, ATLAS:2024jvf}. 
 
To systematically investigate temperature fluctuations in hot QCD matter, we develop a theoretical framework which is general and applicable to temperature fluctuations of arbitrary order. As a specific application, this approach is applied to the QCD thermodynamics described by a 2+1 flavor low energy effective field theory (LEFT) \cite{Wen:2018nkn}, where quantum and thermal fluctuations are encoded self-consistently through the functional renormalization group (fRG). The fRG has proven to be a powerful nonperturbative theoretical method, and is well suited for  studies of properties of the hot QCD matter including the QCD phase diagram, critical end point, and real-time dynamics, etc., \cite{Fu:2019hdw, Braun:2020ada, Braun:2023qak, Tan:2024fuq, Fu:2024rto, Dupuis:2020fhh, Fu:2022gou}.

This paper is organized as follows: In \sec{sec:Wfunction} we first introduce a new thermodynamic state function to characterize the thermodynamics related to the mean transverse momentum fluctuations of charged particles, from which we derive analytic expressions for the temperature fluctuations to arbitrary order in \sec{sec:tem-fluc}. Numerical results are obtained by applying this framework to a 2+1 flavor LEFT within the fRG approach in \sec{sec:numerical}. Our approach demonstrates that temperature fluctuations would be suppressed remarkably as the matter evolves from HRG to QGP with increasing temperature or the baryon chemical potential. Conclusions, discussions and outlook are presented in \sec{sec:conclusion}. In the appendices, we provide some details of the 2+1 flavor LEFT within the fRG approach in \app{app:LEFT2p1}, preliminary results of temperature fluctuations calculated in QCD within fRG in \app{app:cn-QCD}, hyper-order fluctuations of temperature in \app{app:hyper-order}, ratios of temperature fluctuation cumulants in \app{app:cn-ratios}, relation between the temperature $T$ and the mean transverse momentum $\langle p_{T} \rangle$ in \app{app:T-pT-relation}.


%
\begin{figure}[t]
\includegraphics[width=0.45\textwidth]{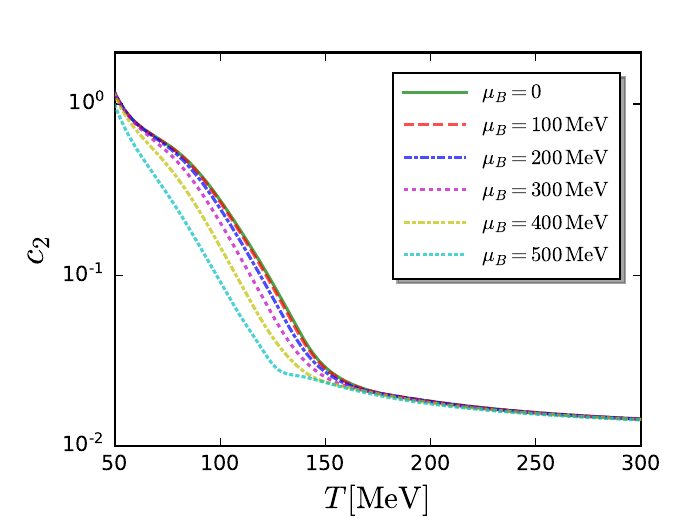}
\caption{Variance of temperature fluctuations as a function of the temperature with several different values of baryon chemical potential.}
\label{fig:c2}
\end{figure}
%

\section{A new thermodynamic state function}
\label{sec:Wfunction}  

We begin with a total derivative of the thermodynamic potential $\Omega$
\begin{align}
    \mathrm{d} \Omega=-S \mathrm{d} T-p \mathrm{d}V-N_B \mathrm{d} \mu_B\,, \label{eq:dOmega}
\end{align}
with the entropy $S$, temperature $T$, pressure $p$, volume $V$, baryon number $N_B$, and the baryon chemical potential $\mu_B$. Although we explicitly show $\mu_B$ as a representative of the conserved charge, Eq.~\labelcref{eq:dOmega} is readily generalized to include additional chemical potentials when other conserved charges are present. The thermodynamic potential $\Omega$ is a state function of $T$, $V$ and $\mu_B$. By implementing a Legendre transformation upon $\Omega$ w.r.t. the conjugate pair $S$ and $T$, we introduce a new state function as
\begin{align}
    W=\Omega+TS\,. \label{eq:W-def}
\end{align}
One immediately recognizes that there is another relation for the state function $W$, that is,
\begin{align}
    W=U-\mu_B N_B\,,\label{}
\end{align}
resulting from general thermodynamical relations, where $U$ denotes the energy. Inserting Eq.~\labelcref{eq:W-def} into Eq.~\labelcref{eq:dOmega}, one arrives at
\begin{align}
    \mathrm{d} W=T\mathrm{d} S-p \mathrm{d}V-N_B \mathrm{d} \mu_B\,, \label{eq:dW}
\end{align}
indicating that $W$ is a state function of $S$, $V$ and $\mu_B$.

In experimental measurements of mean transverse momentum fluctuations, finite acceptance cuts in the rapidity ($y$) and transverse momentum ($p_{T}$) range are applied, which signifies that the system volume in Eq.~\labelcref{eq:dW} is approximately constant. The chemical potential can also be regarded as a constant, which is determined by, e.g., the collision energy, centrality, etc. While this approximation holds for high-energy collisions, we note that $\mu_B$ may vary in low-energy regions, e.g., fixed target collisions at RHIC, due to global baryon number conservation effects \cite{Braun-Munzinger:2020jbk, Vovchenko:2021kxx, Fu:2023lcm}. For the present study, we neglect these corrections and maintain the constant approximation. 

Recent measurements of mean transverse momentum fluctuations at RHIC and the LHC were made at a fixed multiplicity of charged particles $N_{\mathrm{ch}}$, as shown in \cite{ALICE:2023tej, ATLAS:2024jvf, QM2025rutik, QM2025gao}. The $N_{\mathrm{ch}}$ scales directly with the entropy of the system ($N_{\mathrm{ch}}\sim S$), while the resonance decays in the final states or acceptance cuts might modify it slightly. Consequently, the state function $W$ in Eq.~\labelcref{eq:dW} becomes the appropriate thermodynamic potential to describe these experimental observables, since its natural variables correspond directly to the quantities constrained in the measures.

%
\begin{figure*}[t]
\includegraphics[width=0.45\textwidth]{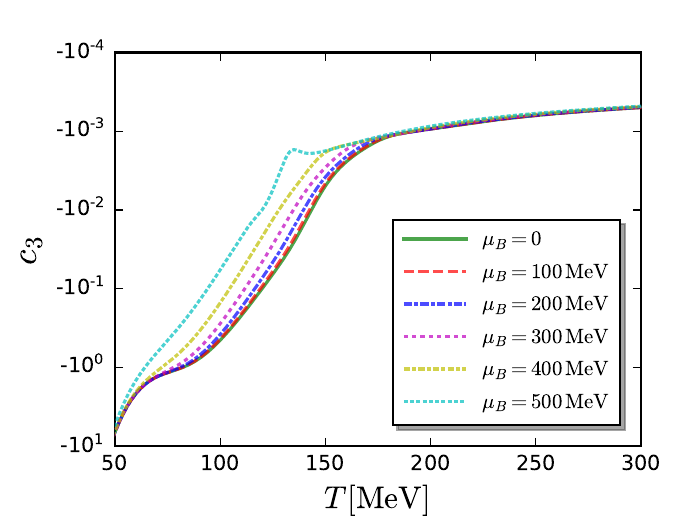}\hspace{0.5cm}
\includegraphics[width=0.45\textwidth]{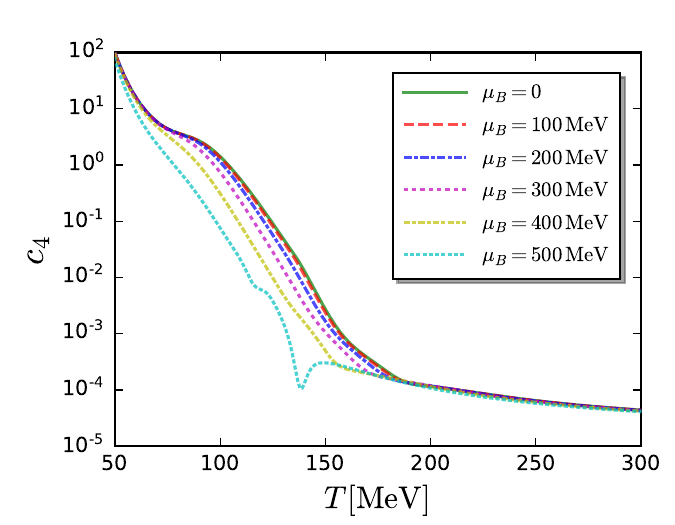}
\caption{Skewness (left panel) and kurtosis (right panel) of temperature fluctuations, i.e., $c_3$ and $c_4$ in Eq.~\labelcref{eq:cn}, as functions of the temperature with several different values of baryon chemical potential.}
\label{fig:c3-4}
\end{figure*}
%

\section{Temperature fluctuations derivations}
\label{sec:tem-fluc}  

Having established the relevance of the state function $W$ in Eq.~\labelcref{eq:dW} for heavy-ion collisions, we now derive the temperature fluctuations, or equivalently, the mean transverse momentum fluctuations of charged particles, computed from the derivative of $W$ w.r.t. $S$ for different orders.

For a fixed volume $V$, we define the intensive quantities: the thermodynamic potential density $w=W/V$ and the entropy density $s=S/V$, one arrives at
\begin{align}
    w=-p+T s\,, \label{}
\end{align}
where $\Omega=-pV$ is used and the entropy density can be obtained from $s=\frac{\partial p}{\partial T}$. The first-order derivative of $w$ w.r.t. $s$ produces the temperature
\begin{align}
    \frac{\partial w}{\partial s}=T\,. \label{}
\end{align}
Then, the $n$-th order fluctuation of temperature is obtained from the $n$-th order derivatives of $w$ w.r.t. $s$, to wit,
\begin{align}
    \langle(\Delta T)^n \rangle=T^{4n-4}\frac{\partial^n w}{\partial s^n}\,, \label{eq:DeltaTn}
\end{align}
with $\Delta T=T-\langle T \rangle$ and $n\geq 2$ ($n \in Z$), where $\langle \cdots \rangle$ denotes the ensemble average. It is convenient to adopt a dimensionless temperature fluctuation
\begin{align}
    c_n=\frac{\langle(\Delta T)^n \rangle}{T^n}\,. \label{eq:cn}
\end{align}
Cumulants $c_n$ can be expressed in terms of the temperature derivatives of pressure through fundamental thermodynamic relations. The first three nontrivial orders corresponding to the variance, skewness, and kurtosis of temperature fluctuations are given by,
\begin{equation}\label{eq:c234}\begin{split}
    & c_2=T^2\left(\frac{\partial^2 p}{\partial T^2}\right)^{-1}\,\\[2ex]
    & c_3=-T^5\left(\frac{\partial^2 p}{\partial T^2}\right)^{-3}\frac{\partial^3 p}{\partial T^3}\,\\[2ex]
    & c_4=T^8\Bigg[3\left(\frac{\partial^2 p}{\partial T^2}\right)^{-5}\left(\frac{\partial^3 p}{\partial T^3}\right)^2-\left(\frac{\partial^2 p}{\partial T^2}\right)^{-4}\frac{\partial^4 p}{\partial T^4}\Bigg]\,.
\end{split} 
\end{equation}
This systematic approach can be extended to higher-order cumulants, e.g., the fifth and sixth hyper-order ones, which are presented in \app{app:hyper-order}.

In relativistic heavy-ion collisions, the temperature fluctuations are measured through EbE mean transverse momentum fluctuations of charged particles, leveraging the approximate linear dependence between the mean transverse momentum and the system temperature, i.e., $\langle p_{T} \rangle = a\,T$ \cite{Gardim:2019xjs, Gardim:2019brr, Giacalone:2020lbm, Gardim:2024zvi}. As explicitly derived for a free relativistic gas in thermal equilibrium with Boltzmann statistics in \app{app:T-pT-relation}, the coefficient is found to be $a=2.356$. In realistic cases such as in experiments, the coefficient might deviate from this value in the ideal case. Therefore, it is useful to construct dimensionless ratios of different orders, thereby eliminating the dependence on the coefficient $a$. The numerical results for those ratios of temperature fluctuation cumulants are presented in \app{app:cn-ratios}.

Note that results of hydrodynamic simulations in \cite{Gardim:2019xjs} indicate that while the $\langle p_{T} \rangle$ and the effective $T$ are sensitive to details of the hydrodynamic modeling, e.g., the initial density profile, transport coefficients, equation of state, etc., the model dependence disappears when considering the ratio $\langle p_{T} \rangle/T$, i.e., the proportionality constant $a$. Moreover, it is also found that $a$ is a constant across different centrality and different collision energy from RHIC to LHC. The hydrodynamic studies motivate the assumption that the almost constant $a$ also holds in event-by-event collisions, since every event is independent of each other. However, this assumption should be verified in the event-by-event simulations of dynamical models in the near future.

\section{Numerical results}
\label{sec:numerical} 

%
\begin{figure}[t]
\includegraphics[width=0.48\textwidth]{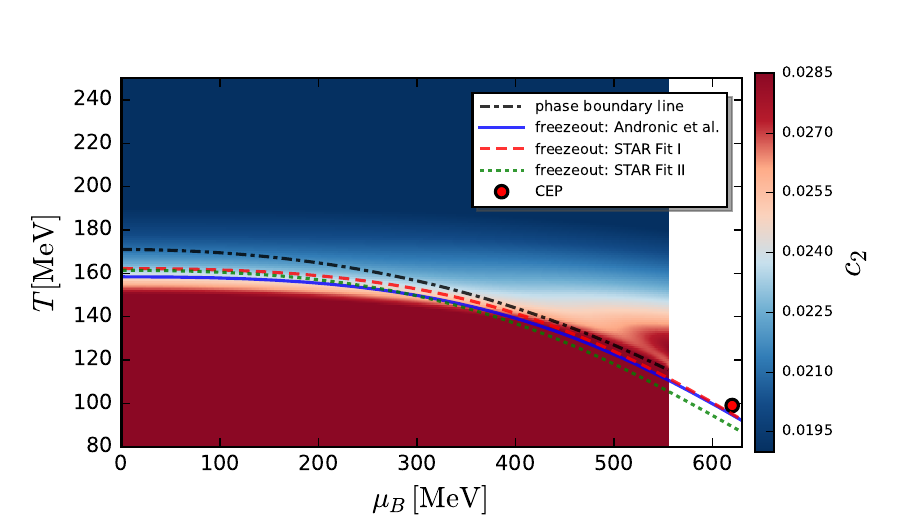}
\caption{Heatmap of the variance of temperature fluctuations in the QCD phase diagram. The CEP and phase boundary line obtained in the LEFT, and three representative freeze-out curves~\cite{Andronic:2017pug, Fu:2021oaw, Fu:2023lcm} are also depicted.}
\label{fig:c2-phase}
\end{figure}
%

%
\begin{figure*}[t]
\includegraphics[width=0.48\textwidth]{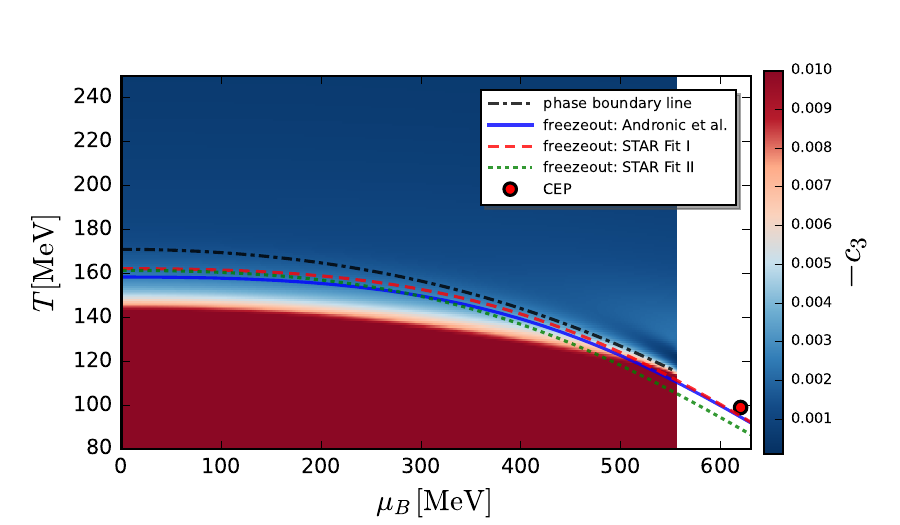}\hspace{0.2cm}
\includegraphics[width=0.48\textwidth]{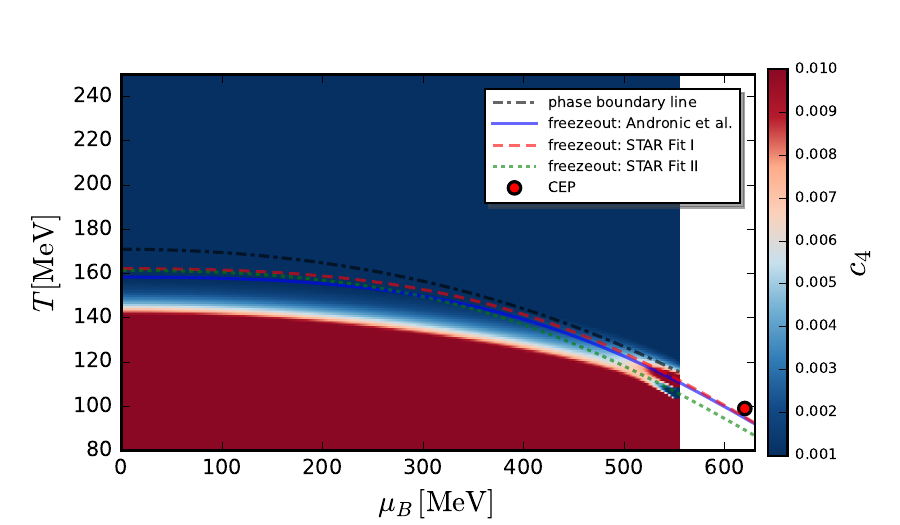}
\caption{Heatmaps of the skewness (left panel) and kurtosis (right panel) of temperature fluctuations in the QCD phase diagram. The CEP and phase boundary line obtained in the LEFT, and three representative freeze-out curves~\cite{Andronic:2017pug, Fu:2021oaw, Fu:2023lcm} are also depicted. Note that here $-c_3$ instead of $c_3$ is depicted.}
\label{fig:c3-4-phase}
\end{figure*}
%

%
\begin{figure*}[t]
\includegraphics[width=0.33\textwidth]{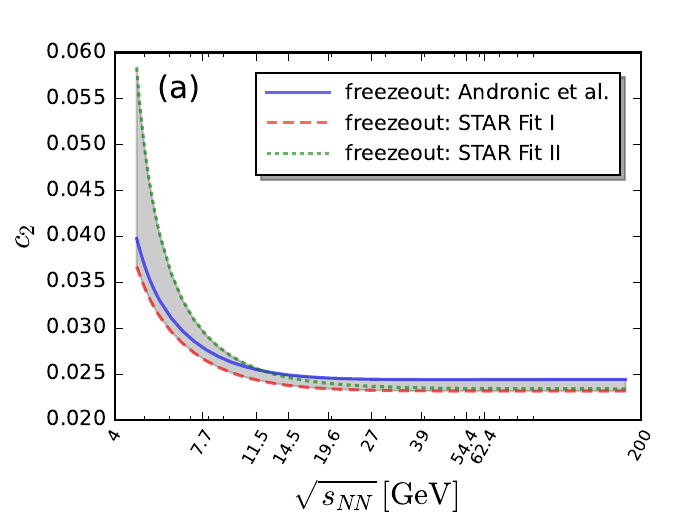}
\includegraphics[width=0.33\textwidth]{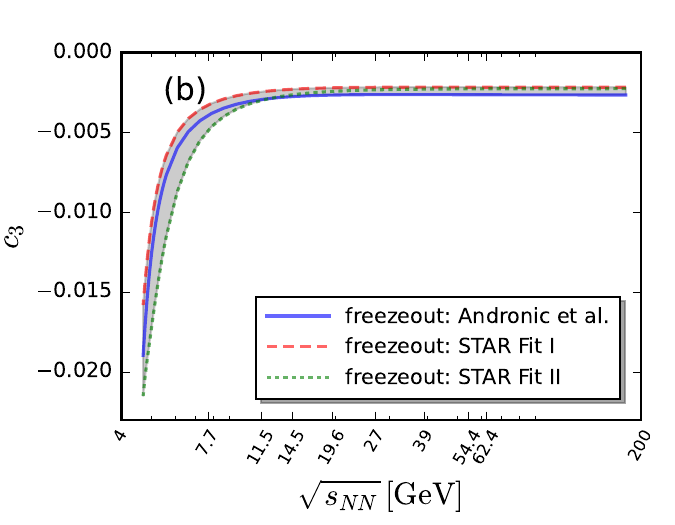}
\includegraphics[width=0.33\textwidth]{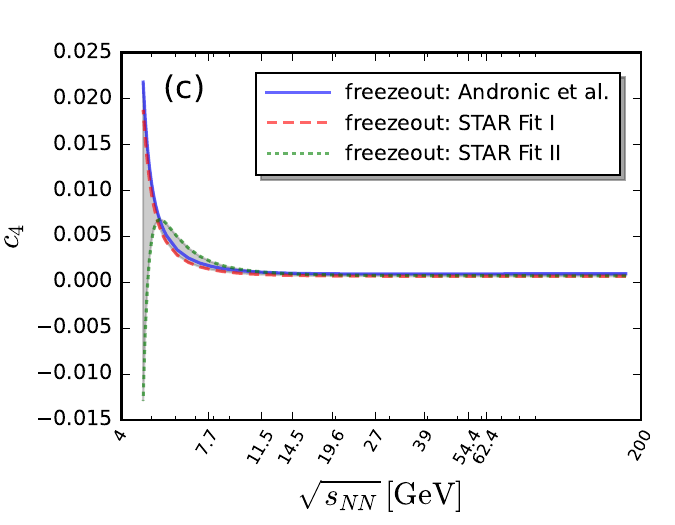}
\caption{Variance (panel a), skewness (panel b), and kurtosis (panel c) of temperature fluctuations, i.e., $c_2$, $c_3$ and $c_4$, as functions of the collision energy, calculated on the three representative freeze-out curves~\cite{Andronic:2017pug, Fu:2021oaw, Fu:2023lcm}.}
\label{fig:cnfreezeout}
\end{figure*}
%

We investigate QCD thermodynamics employing a 2+1 flavor LEFT within the fRG approach. As demonstrated in Ref. \cite{Wen:2018nkn}, this approach yields an equation of state (EoS) and baryon number fluctuations consistent with lattice QCD calculations. The setup of our LEFT is also recapitulated in \app{app:LEFT2p1}.

To proceed, we systematically calculate the temperature derivatives of pressure: 
\begin{align}
    \chi_n=T^{n-4}\frac{\partial^n p}{\partial T^n}\,, \label{eq:chi}
\end{align}
which is dimensionless by normalization with appropriate powers of $T$. From the state function $\Omega$ in Eq.~\labelcref{eq:dOmega}, we identify the first and second order derivatives, $\chi_1$ and $\chi_2$, are just related to the entropy and heat capacity, respectively. Higher-order $\chi_n$ ($n \geq 2$) can be interpreted as entropy fluctuations of different orders. The numerical results of $\chi_n$ from the first to sixth orders calculated in the 2+1 flavor LEFT-fRG framework are presented in \Fig{fig:chi1-2}, \Fig{fig:chi3-4}, and \Fig{fig:chi5-6} in \app{app:LEFT2p1}. We found that the entropy fluctuations increase and oscillate near the chiral crossover, and the strength and amplitude of the oscillation increase with the order of fluctuations or the value of the baryon chemical potential. 

The temperature fluctuations in Eq.~\labelcref{eq:cn} can be reformulated in terms of $\chi_n$, defined in Eq.~\labelcref{eq:chi}. For the lowest-order cumulants, we obtain
\begin{align}
    c_2=&\frac{1}{\chi_2}\,, \qquad c_3=-\frac{\chi_3}{{\chi_2}^3}\,, \qquad c_4=3\frac{{\chi_3}^2}{{\chi_2}^5}-\frac{\chi_4}{{\chi_2}^4}\,. \label{eq:cn-chin}
\end{align}

The variance of temperature fluctuations, $c_2$, is inversely proportional to the variance of entropy fluctuations, i.e., the heat capacity $\chi_2$, as demonstrated in Fig.~\ref{fig:c2}. We observe that $c_2$ decreases with increasing temperature, reflecting the opposite trend of $\chi_2$ as shown in the right panel of \Fig{fig:chi1-2}. This behavior indicates a significant suppression of temperature fluctuations in QGP phase compared to those in HRG phase. The suppression is more remarkable for high-order temperature fluctuations, as is evident in Fig.~\ref{fig:c3-4} 
(note the logarithmic $y$-axis). A direct consequence of the suppression of temperature fluctuations at high temperature is that the distribution of temperature is wider in the region of lower temperature, which implies a negative skewness, as confirmed in the left panel of Fig.~\ref{fig:c3-4}. While the kurtosis remains positive in most cases, its sign may reverse near the chiral crossover as it is sharpened continuously with the increase in the baryon chemical potential. The numerical results of the hyper-order $c_5$ and $c_6$ cumulants are presented in Fig.~\ref{fig:c5-6} in \app{app:hyper-order}. The temperature fluctuations obtained here in the LEFT are also compared with the preliminary results obtained in QCD within the fRG in \Fig{fig:c1-2-QCD} in \app{app:cn-QCD}, where qualitative agreement is observed though there are some quantitative discrepancies.

We compute $c_2$, $c_3$, $c_4$ on the QCD phase diagram in \Fig{fig:c2-phase} and \Fig{fig:c3-4-phase}, together with the phase boundary line and the critical end point obtained in the LEFT, located at 
\begin{align}
    (T_{_\mathrm{CEP}},\mu_{B_{\mathrm{CEP}}})=(99, 620)\,\mathrm{MeV}\,, \label{eq:CEP}
\end{align}
which is qualitatively consistent with the recent results obtained from functional QCD, see, e.g., \cite{Fu:2019hdw, Gao:2020fbl, Gunkel:2021oya}. Moreover, on the phase diagram we also depict three representative freeze-out curves, which were discussed in detail in \cite{Fu:2021oaw} and also used in \cite{Fu:2023lcm}. The freeze-out curve of Andronic {\it et al.} is obtained by parameterizing the freeze-out data from \cite{Andronic:2017pug}. The other two freeze-out curves (STAR Fit I and II) are based on the freeze-out data from the STAR experiment \cite{Adamczyk:2017iwn}, where all the freeze-out data from STAR are used for the former, and for the latter, some flawed data of collision energy are dropped from general considerations, see \cite{Fu:2021oaw} for more details.

Note that the calculations of temperature fluctuations in \Fig{fig:c2-phase} and \Fig{fig:c3-4-phase} are performed up to $\mu_B \sim 550$ MeV. Results of higher $\mu_B$ are not accessible for the moment due to the significantly increasing difficulty in this region. Nevertheless, this does not prevent us from gaining qualitative insight into the structure of the temperature fluctuations on the plane of phase diagram. Obviously, the magnitude of temperature fluctuations below the phase boundary is larger than that above the phase boundary. With the increase of $\mu_B$ as it approaches the CEP, there are some nontrivial structures developed near the phase boundary. This is obvious for the high-order fluctuations, e.g., in the plane of $c_4$, some red and blue bands appear in the proximity of CEP.

We further compute the temperature fluctuations on the chemical freeze-out curves, which allow us to investigate their dependence on the collision energy. The relevant results are shown in \Fig{fig:cnfreezeout}. Results on the three different freeze-out curves are employed to investigate the errors arising from the uncertainty in the determination of freeze-out curves. From Fig.~\ref{fig:cnfreezeout} one observes that with the decrease of collision energy from high to low, all the temperature fluctuations investigated here, $c_2$, $c_3$ and $c_4$, exhibit mild dependence on the collision energy firstly. When the collision energy is reduced below approximately $14.5$ GeV, the magnitude of $c_2$ and $c_3$ increases significantly due to the increasingly sharp chiral crossover with the increase of $\mu_B$, while the skewness $c_3$ remains negative. Due to the nontrivial structure near the CEP discussed above, the kurtosis $c_4$ in the regime of low collision energy is highly sensitive to the choice of freeze-out curves, resulting in uncertainties too large to draw a definitive conclusion.

There is a caveat when the results in \Fig{fig:cnfreezeout} are compared with experimental data. As we have discussed above, the results in \Fig{fig:cnfreezeout} are obtained by evaluating the temperature fluctuations in thermal equilibrium at the chemical freeze-out. To facilitate a better comparison with experiments, it is desirable to include other effects, such as the initial state geometry fluctuations,  flow fluctuations, etc., in a dynamical evolution. This is beyond the scope of this paper, which will be investigated in the future.

\section{Conclusions, discussions and outlook}
\label{sec:conclusion} 

We have studied temperature fluctuations in hot QCD matter through a newly introduced thermodynamic state function that potentially connects to the mean transverse momentum fluctuations measured in heavy-ion collisions. Our approach yields analytic expressions for arbitrary-order temperature fluctuations, revealing their fundamental relationship with entropy, heat capacity, and high-order entropy fluctuations. Implementing this general formalism in a 2+1 flavor LEFT within the fRG, we obtain numerical results that quantify the temperature fluctuations across different thermodynamic regimes for the first time.

As the system transitions from the HRG to QGP phase with the increasing temperature or baryon chemical potential, the heat capacity of QCD matter increases substantially. This implies that a tiny change of the temperature would cost a huge amount of energy in the regime of high temperature. Therefore, the temperature are therefore reduced in comparison to the case in the regime of low temperature. In other words, the temperature fluctuations would be suppressed remarkably as the matter evolves from the HRG phase to the QGP phase with the increase in temperature or baryon chemical potential, as demonstrated in our calculations. The fact that temperature fluctuations at high temperature are smaller than those at low temperature leads to another direct consequence, i.e., a negative skewness of temperature fluctuations. Such a signature emerges because the increasingly narrow fluctuation distribution at high temperature creates an asymmetric probability density weighted toward lower temperatures. 

Note that these findings are general and model independent. They arise from the fact that the heat capacity of the QCD matter increases significantly from the HRG phase to the QGP phase. In the meantime, they provide a candidate observable for discovering the thermodynamic temperature fluctuations in upcoming heavy-ion collision experiments at e.g., RHIC-BES, FAIR-CBM, NICA, and HIAF, which also paves a novel way to study QCD thermodynamics and QCD phase diagram through measurements of the mean transverse momentum fluctuations of charged particles.

During the review of this manuscript, the STAR Collaboration reported the measurements of the variance of the mean $p_T$ fluctuations of charged particles across different collision energy \cite{STAR:2026vjv}, where a non-monotonic dependence on the collision energy was observed in central collisions. While a quantitative comparison between the theoretical calculations presented here and the experimental data remains challenging, we note that the theoretical curve in the left panel of \Fig{fig:cnfreezeout} is qualitatively consistent with the data. We anticipate future measurements of higher-order mean $p_T$ fluctuations, in particular the skewness of the mean $p_T$ distributions.

Here we discuss the relations between temperature fluctuations and the conventional net-proton fluctuations, which are complementary observables. In the QCD phase diagram in terms of the temperature and baryon chemical potential, the net-proton fluctuations are related to the direction of the baryon chemical potential, while the temperature fluctuations probe the temperature direction. A comprehensive study of the QCD criticality along both directions therefore provides complementary constraints on the two-dimensional structure of the QCD phase diagram. Moreover, using the net-proton fluctuations to search for the CEP relies on the coupling between the conserved charges, e.g., the baryon number, and the critical mode, i.e., the sigma mode of CEP, while the temperature fluctuation does not rely on such coupling, and it represents a complementary observable sensitive to critical thermodynamics. Finally, partly due to the aforementioned reason, the critical exponents of temperature fluctuations are the well-known universal exponents. For example the variance of the temperature fluctuations equals the inverse of the heat capacity as shown in \Eq{eq:cn-chin}. However, the critical exponents of baryon or proton fluctuations are not well determined, and assumptions or models are required in the computation of these exponents. In the future, should a signal of the CEP be observed in the net-proton fluctuation experiments, an independent confirmation from complementary observables such as temperature fluctuations would provide valuable corroborating evidence.

\section*{Acknowledgements}

We thank Fei Gao,  Xuguang Huang, Jan M. Pawlowski for discussions and comments. W.J. Fu and S.Yin also would like to thank the members of the fQCD collaboration \cite{fQCD}, especially Rui Wen, for collaborations on related projects.
J.H. Chen\ is supported by the National Key Research and Development Program of China under Contract No. 2022YFA1604900, by the National Natural Science Foundation of China under Contract No. 12025501.
W.J. Fu\ is supported by the National Natural Science Foundation of China under Contract Nos.\ 12447102, 12175030. 
S. Yin\ is supported by the Alexander v.\ Humboldt Foundation. 
S. Yin\ acknowledges support by the Deutsche Forschungsgemeinschaft (DFG, German Research Foundation) through the CRC-TR 211 ``Strong-interaction matter under extreme conditions"– project number 315477589 – TRR 211.
C. Zhang\ is supported by the National Key Research and Development Program of China under Contract No. 2024YFA1612600, National Natural Science Foundation of China under Contract No. 12547102, Shanghai Pujiang Talents Program under Contract No. 24PJA009.

\appendix


\section{2+1 flavor low energy effective field theory within the fRG}
\label{app:LEFT2p1}

In this appendix, we recapitulate the setup of the 2+1 flavor low energy effective field theory (LEFT) within the functional renormalization group used in this work. More details about the 2+1 LEFT can be found in \cite{Wen:2018nkn}. The effective action reads
\begin{align}
    \Gamma_{k}[\Phi]&=\int_x \bigg\{ \bar{q} \big[\gamma_\mu \partial_\mu-\gamma_0(\mu+igA_0)\big]q+h_k\,\bar{q} \,\Sigma_5 q\nonumber\\[2ex]
    &\quad+\text{tr}\big(\partial_\mu \Sigma \cdot \partial_\mu\Sigma^\dagger \big)+V_{\mathrm{\tiny{matt}},k}(\phi)+V_{\mathrm{\tiny{glue}}}(L, \bar L)\bigg\}\,,\label{eq:action}
\end{align}
with the shorthand notation $\int_{x}=\int_0^{\beta}d x_0 \int d^3 x$ and $\beta=1/T$, where $T$ stands for the temperature. The subscript $k$ in $\Gamma_{k}$ indicates that an infrared (IR) cutoff is applied to the effective action, such that quantum and thermal fluctuations of momenta $p\lesssim k$ are suppressed. The full effective action is resolved as $k \to 0$, and thus $k$ plays a role as the renormalization group (RG) scale. The field $\Phi=(q, \bar q, \phi)$ includes the three-flavor quark field $q=(q_u, q_d, q_s)^\intercal$ and the scalar and pseudoscalar meson fields $\phi=(\sigma, \pi)$. The mesons are in the adjoint representation of the $\mathrm{U}(N_f=3)$ group, which reads
\begin{align}
    \Sigma=T^a(\sigma^a+i \pi^a)\,, \quad a=0,\,1,...,8\,,\label{}
\end{align}
with
\begin{align}
    T^{0}=\frac{1}{\sqrt{2N_{f}}}\mathbb{1}_{N_{f}\times N_{f}}\,,\label{}
\end{align}
and
\begin{align}
    T^a=\frac{\lambda^a}{2} \quad(a=1,...,8)\,, \label{}
\end{align}
where $\lambda^a$ are the Gell-Mann matrices. The quark and meson fields interact with each other through the Yukawa coupling $h_k$ with
\begin{align}
    \Sigma_5=T^a(\sigma^a+i \gamma_5\pi^a)\,.\label{}
\end{align}
The quark chemical potential $\mu$ in \labelcref{eq:action} is related to the baryon chemical potential $\mu_B$ with $\mu=\mu_B/3$, where other chemical potentials, e.g., the chemical potentials for the electric charge and strangeness, are assumed to be vanishing.

%
\begin{figure}[t]
\includegraphics[width=0.45\textwidth]{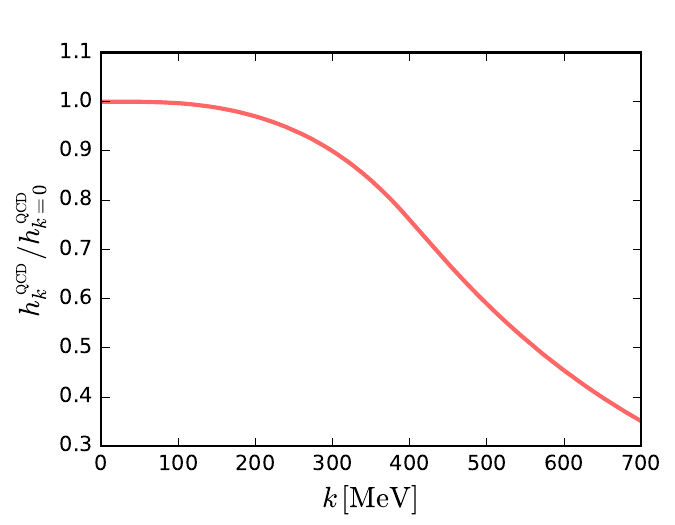}
\caption{Yukawa coupling, normalized to unity for $k=0$, as a function of the RG scale $k$, computed from the first-principles QCD within the fRG in the vacuum \cite{Fu:2019hdw}.}
\label{fig:hQCD-k}
\end{figure}
%

%
\begin{figure*}[t]
\includegraphics[width=0.45\textwidth]{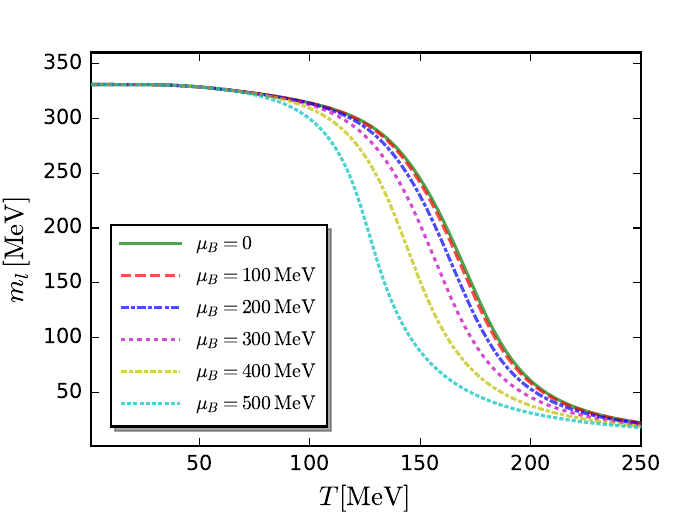}\hspace{0.5cm}
\includegraphics[width=0.45\textwidth]{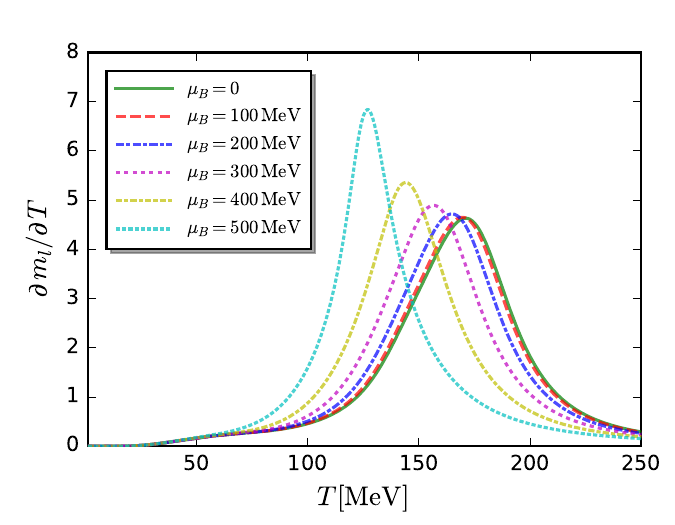}
\caption{Constituent light quark mass (left panel) and its derivative to the temperature (right panel) as functions of the temperature with several values of baryon chemical potential.}
\label{fig:ml}
\end{figure*}
%

%
\begin{figure}[t]
\includegraphics[width=0.45\textwidth]{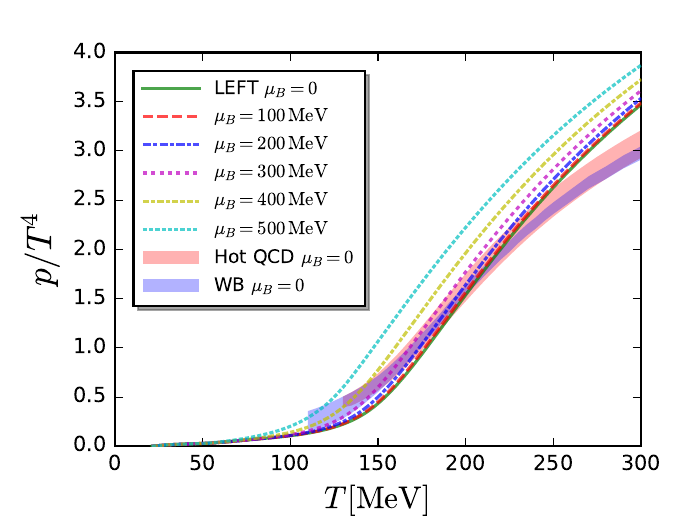}
\caption{Pressure normalized by $T^4$ as a function of the temperature with several different values of $\mu_B$ obtained in the 2+1 flavor LEFT, in comparison to the results obtained in lattice QCD at vanishing $\mu_B$ from the HotQCD collaboration \cite{HotQCD:2014kol} and the Wuppertal-Budapest collaboration (WB) \cite{Borsanyi:2013bia}.}
\label{fig:pressure}
\end{figure}
%

%
\begin{figure*}[t]
\includegraphics[width=0.45\textwidth]{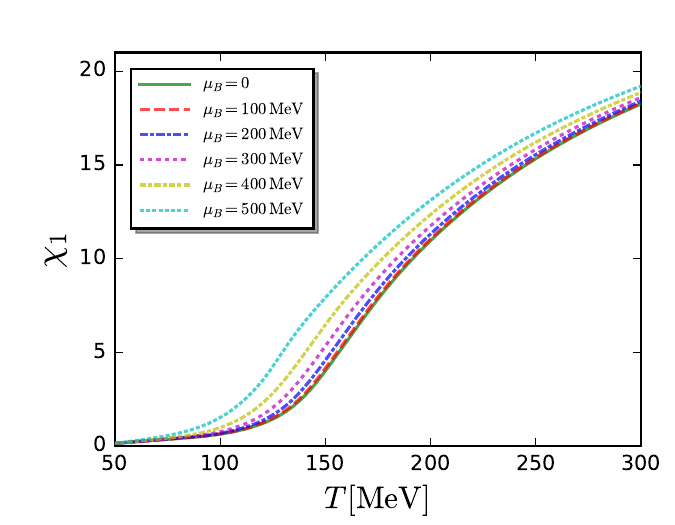}\hspace{0.5cm}
\includegraphics[width=0.45\textwidth]{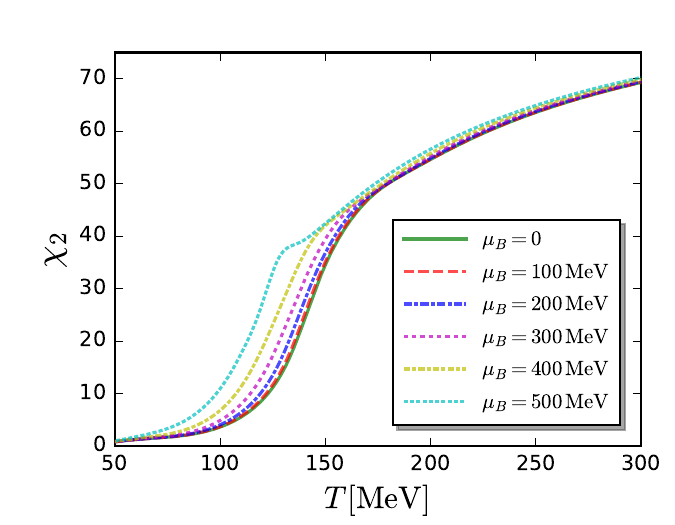}
\caption{Dimensionless entropy (left panel) and heat capacity (right panel) normalized by appropriate powers of $T$, i.e., $\chi_1$ and $\chi_2$, as functions of the temperature with several values of baryon chemical potential.}
\label{fig:chi1-2}
\end{figure*}
%

%
\begin{figure*}[t]
\includegraphics[width=0.45\textwidth]{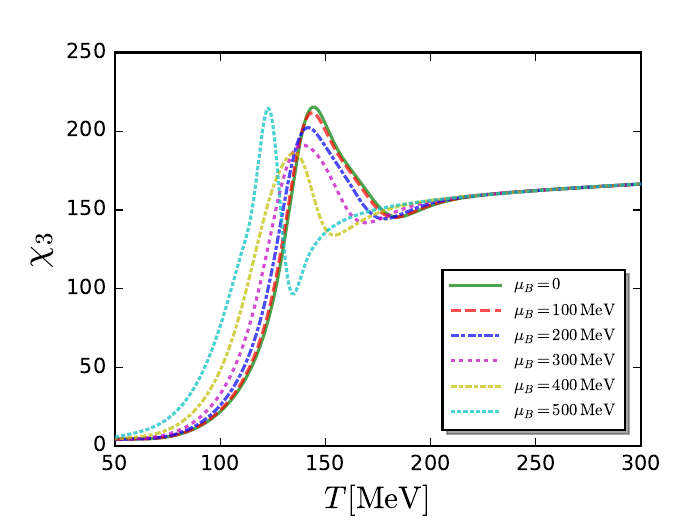}\hspace{0.5cm}
\includegraphics[width=0.45\textwidth]{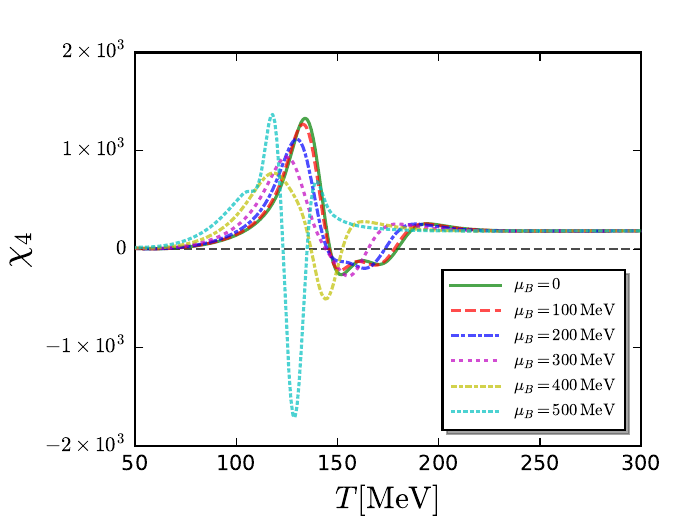}
\caption{Skewness (left panel) and kurtosis (right panel) of the entropy fluctuations, i.e., $\chi_3$ and $\chi_4$, as functions of the temperature with several values of baryon chemical potential.}
\label{fig:chi3-4}
\end{figure*}
%

%
\begin{figure*}[t]
\includegraphics[width=0.45\textwidth]{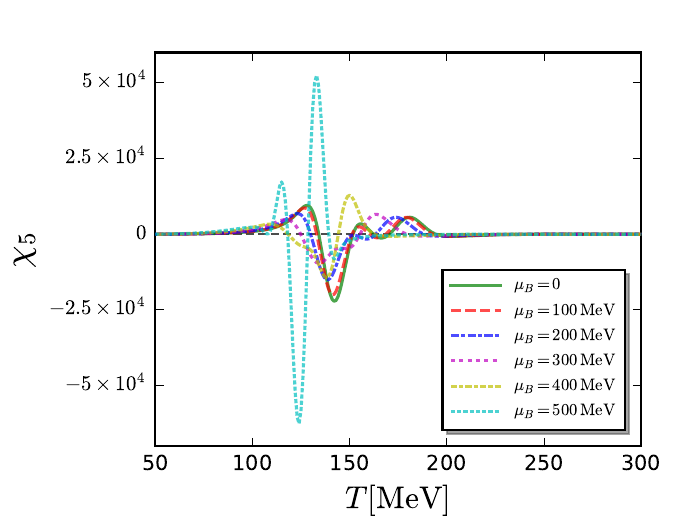}\hspace{0.5cm}
\includegraphics[width=0.45\textwidth]{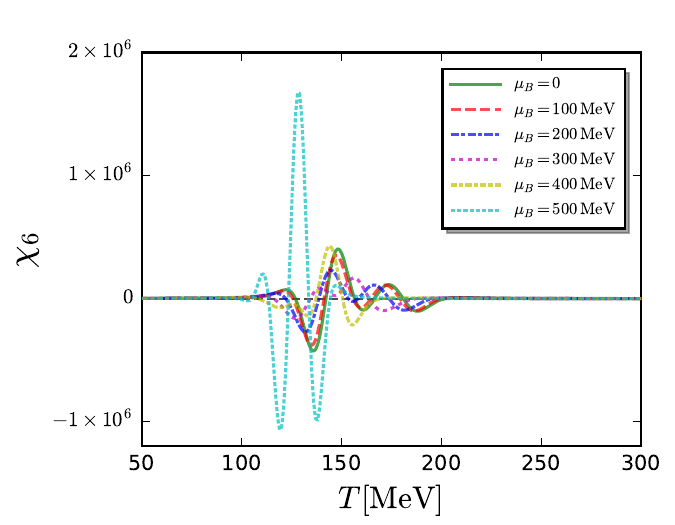}
\caption{Fifth (left panel) and sixth (right panel) order fluctuations of the entropy, i.e., $\chi_5$ and $\chi_6$, as functions of the temperature with several values of baryon chemical potential.}
\label{fig:chi5-6}
\end{figure*}
%

The mesonic potential in \labelcref{eq:action}, i.e., the matter sector of the effective potential, reads
\begin{align}
    V_{\text{\tiny{matt}},k}(\phi)&=\tilde{V}_k(\rho_1,\rho_2)-c_A \xi-c_l\sigma_l-c_s\sigma_s\,, \label{eq:Vmatt}
\end{align}
with
\begin{align}
    \rho_1&=\text{tr}(\Sigma \cdot \Sigma^\dagger)\,, \label{eq:rho1}\\[2ex]
    \rho_2&=\text{tr}\Big(\Sigma \cdot \Sigma^\dagger-\frac{1}{3}\,\rho_1\,\mathbb{1}_{3\times 3}\Big)^2 \,,\label{eq:rho2}
\end{align}
where $\rho_1$ and $\rho_2$ are invariant under the transformations $\mathrm{SU}_{\mathrm{V}}(3)\times \mathrm{SU}_{\mathrm{A}}(3)\times \mathrm{U}_{\mathrm{V}}(1)\times \mathrm{U}_{\mathrm{A}}(1)$ in the flavor space. Here $\sigma_l$ and $\sigma_s$ indicate the scalar mesons of light and strange quarks, respectively. The relevant strength constants $c_l$ and $c_s$ result in explicit breaking of the chiral symmetry, as well as the breaking of the flavor symmetry from the three-flavor case to that of 2+1 flavors. The $\mathrm{U}_{\mathrm{A}}(1)$ symmetry is broken by the Kobayashi-Maskawa-'t Hooft determinant, viz.,
\begin{align}
    \xi=\det(\Sigma)+\det(\Sigma^\dagger)\,, \label{}
\end{align}
arising from quantum fluctuations, whose strength is controlled by the constant $c_A$.

The glue dynamics is encoded in the glue potential $V_{\mathrm{\tiny{glue}}}$ in \labelcref{eq:action}, also known as the Polyakov loop potential. The Polyakov loop is related to the temporal gluon background field $A_0$, that reads
\begin{align}
  L(\bm{x})=\frac{1}{N_c} \left\langle \Tr\, {\cal P}(\bm x)\right\rangle \,,\quad  \bar L (\bm{x})=\frac{1}{N_c} \langle
  \Tr\,{\cal P}^{\dagger}(\bm x)\rangle \,,\label{eq:Lloop}
\end{align}
with 
\begin{align}
  {\cal P}(\bm x)= \mathcal{P}\exp\Big(ig\int_0^{\beta}d\tau A_0(\bm{x},\tau)\Big)\,,\label{eq:Ploop}
\end{align}
where $\mathcal{P}$ on the right side denotes the path ordering, and $g$ is the strong coupling constant.

In this work, we employ the Haar glue potential \cite{Lo:2013hla, Wen:2018nkn}, 
\begin{align}
    V_{\text{\tiny{glue}}}(L, \bar L)=T^4 \,\bar V_{\text{glue-Haar}}\,,\label{eq:Vglue}
\end{align}
with
\begin{align}
    \bar V_{\text{glue-Haar}} &= -\frac{\bar a(T)}{2} \bar L L + \bar b(T)\ln M_H(L,\bar{L})\nonumber\\[2ex]
    &\quad+ \frac{\bar c(T)}{2} (L^3+\bar L^3) + \bar d(T) (\bar{L} L)^2\,,\label{eq:GluepHaar}
\end{align}
where the Haar measure reads
\begin{align}
    M_H (L, \bar{L})&= 1 -6 \bar{L}L + 4 (L^3+\bar{L}^3) - 3  (\bar{L}L)^2\,.
\end{align}
The temperature dependence of the coefficients $\bar a$, $\bar c$, $\bar d$ in \labelcref{eq:GluepHaar} is parameterized as
\begin{align}
    x(T) &= \frac{x_1 + x_2/(t+1) + x_3/(t+1)^2}{1 + x_4/(t+1) + x_5/(t+1)^2}\,,\quad x\in (\bar a, \bar c, \bar d)\,,\label{eq:xT}
\end{align}
and that of $\bar b$ as
\begin{align}
    \bar b(T) &=\bar b_1 (t+1)^{-\bar b_4}\left (1 -e^{\bar b_2/(t+1)^{\bar b_3}} \right)\,.\label{eq:bT}
\end{align}
Here $t$ in \labelcref{eq:xT,eq:bT} is the reduced temperature, that is
\begin{align}
    t&=\alpha (T-T_c^\mathrm{\tiny{glue}})/T_c^\mathrm{\tiny{glue}}\,,\label{}
\end{align}
where the parameters $T_c^\mathrm{\tiny{glue}}=230$ MeV and $\alpha=0.5$ are used throughout this work. The values of other parameters in \labelcref{eq:xT,eq:bT} can be found in \cite{Lo:2013hla, Wen:2018nkn}.

Moreover, using the same method in \cite{Fu:2023lcm}, we employ the dependence of the Yukawa coupling on the RG scale $k$ calculated from the first-principles QCD \cite{Fu:2019hdw}, as an input for the LEFT. Then the Yukawa coupling in the LEFT now reads
\begin{align}
    h_k&=h_0 \frac{h_{k}^{\mathrm{QCD}}}{h_{k=0}^{\mathrm{QCD}}}\,,\label{eq:hk}
\end{align}
where $h_{k}^{\mathrm{QCD}}$ is computed by using the fRG approach to the first-principles QCD in the vacuum \cite{Fu:2019hdw}, as shown in Fig.~\ref{fig:hQCD-k}. Here the parameter in the LEFT $h_0=14$ is determined by fitting the constituent light $u$ and $d$ quark mass $m_l=330$ MeV. In the meantime, it produces the constituent strange $s$ quark mass $m_s=568$ MeV.  Furthermore, we use the same values of parameters in the matter sector in \labelcref{eq:Vmatt} as those in \cite{Wen:2018nkn}.

In the left panel of Fig.~\ref{fig:ml}, we show the constituent masses for the $u$, $d$ light quarks calculated in the 2+1 flavor LEFT, depicted as functions of the temperature at several values of $\mu_B$. Their respective derivatives with respect to the temperature are shown in the right panel of Fig.~\ref{fig:ml}, from which one can determine the pseudo-critical temperature for the chiral crossover through the location of the peak. Figure~\ref{fig:pressure} displays the temperature dependence of pressure calculated in our 2+1 flavor LEFT-fRG framework for baryon chemical potential $\mu_B$ ranging from $\mu_B=0$ to 500 MeV. The relevant lattice results at $\mu_B=0$ are also presented for comparison \cite{HotQCD:2014kol, Borsanyi:2013bia}. One can see that the LEFT is consistent with the lattice when $T \lesssim 230$ MeV, which is the temperature range most concerned in this work, while the LEFT overshoots a bit the lattice results in the region of high temperature. The temperature derivatives of pressure,  from the first to sixth orders, are presented in Figs.~\ref{fig:chi1-2} to~\ref{fig:chi5-6}, which stand for the entropy and its fluctuations of different orders.

\section{Preliminary results of temperature fluctuations calculated in QCD within fRG}
\label{app:cn-QCD}

%
\begin{figure*}[t]
\includegraphics[width=0.45\textwidth]{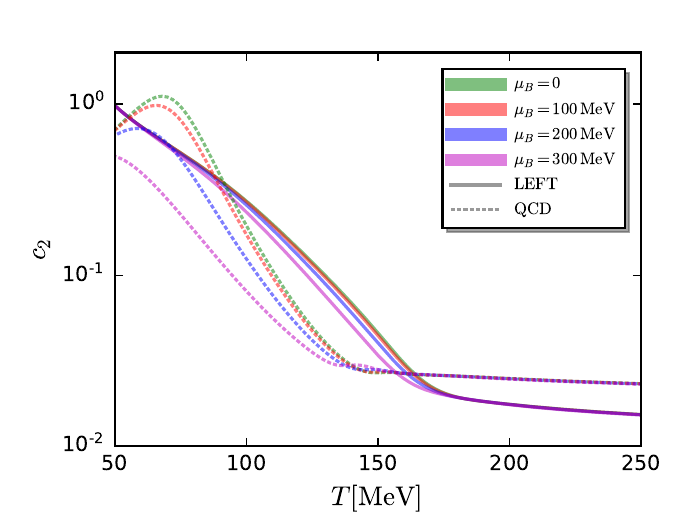}\hspace{0.5cm}
\includegraphics[width=0.45\textwidth]{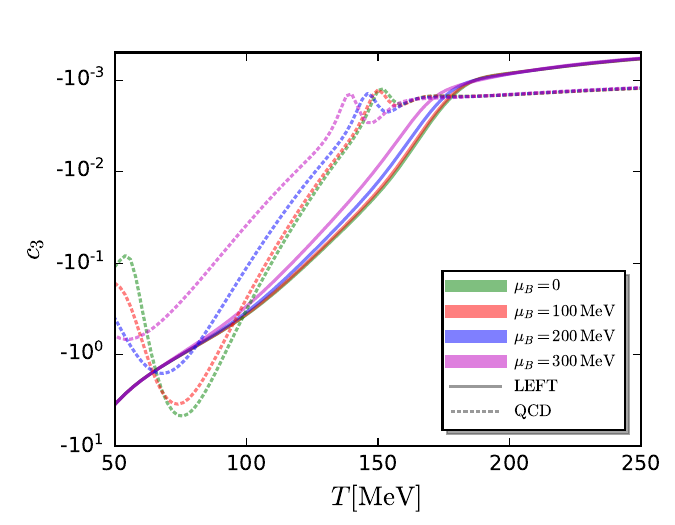}
\caption{Comparison between the LEFT and QCD computations for the second- and third-order fluctuations of temperature, i.e., $c_2$ (left panel) and $c_3$ (right panel), as functions of the temperature with several values of baryon chemical potential. The LEFT and QCD results are denoted by the solid and dashed lines, respectively.}
\label{fig:c1-2-QCD}
\end{figure*}
%

In addition, we also compare the temperature fluctuations obtained in the LEFT with the preliminary results obtained in QCD within the fRG, whose setup is detailed in \cite{Fu:2019hdw}. The relevant results are presented in Fig.~\ref{fig:c1-2-QCD}. One can see that although there are some differences, the variance and skewness of temperature fluctuations calculated from the LEFT and QCD are consistent with each other qualitatively. Note that the main conclusions obtained from the calculations in LEFT are corroborated by the QCD results, e.g., the skewness of temperature fluctuations is negative and the variance decreases significantly with the increase of temperature.

\section{Hyper-order fluctuations of temperature}
\label{app:hyper-order}

%
\begin{figure*}[t]
\includegraphics[width=0.45\textwidth]{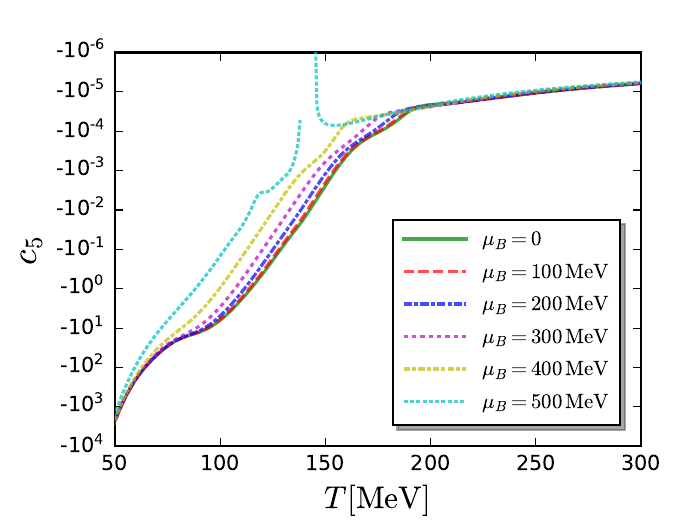}\hspace{0.5cm}
\includegraphics[width=0.45\textwidth]{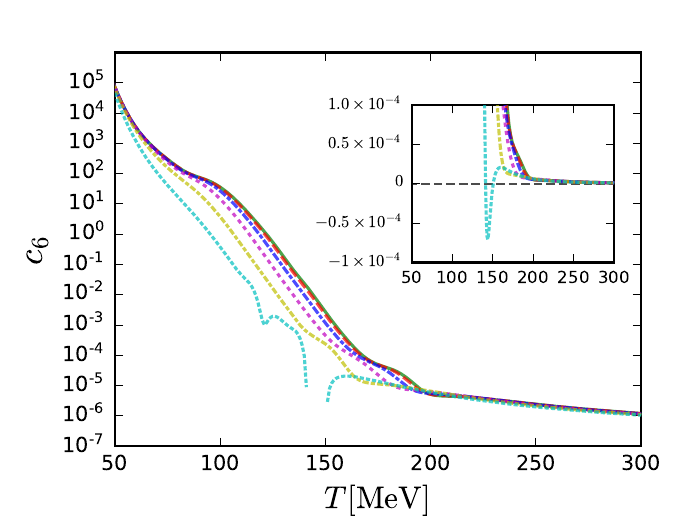}
\caption{Hyper-order fluctuations of temperature, $c_5$ (left panel) and $c_6$ (right panel), as functions of the temperature with several different values of baryon chemical potential. The inset on the right shows the plot by using the linear $y$-axis, where the zero-crossing for $c_6$ at high baryon chemical potential is clear.}
\label{fig:c5-6}
\end{figure*}
%

The fifth- and sixth-order fluctuations of temperature are given by
\begin{align}
    c_5&=T^{11}\Bigg[-15\left(\frac{\partial^2 p}{\partial T^2}\right)^{-7}\left(\frac{\partial^3 p}{\partial T^3}\right)^3-\left(\frac{\partial^2 p}{\partial T^2}\right)^{-5}\frac{\partial^5 p}{\partial T^5}\nonumber\\[2ex]
    &\quad+10\left(\frac{\partial^2 p}{\partial T^2}\right)^{-6}\frac{\partial^3 p}{\partial T^3}\frac{\partial^4 p}{\partial T^4}\Bigg]\,,\label{eq:c5}\\[2ex]
    c_6&=T^{14}\Bigg[105\left(\frac{\partial^2 p}{\partial T^2}\right)^{-9}\left(\frac{\partial^3 p}{\partial T^3}\right)^4-105\left(\frac{\partial^2 p}{\partial T^2}\right)^{-8}\nonumber\\[2ex]
    &\quad\times\left(\frac{\partial^3 p}{\partial T^3}\right)^2\frac{\partial^4 p}{\partial T^4}+10\left(\frac{\partial^2 p}{\partial T^2}\right)^{-7}\left(\frac{\partial^4 p}{\partial T^4}\right)^2\nonumber\\[2ex]
    &\quad+15\left(\frac{\partial^2 p}{\partial T^2}\right)^{-7}\frac{\partial^3 p}{\partial T^3}\frac{\partial^5 p}{\partial T^5}-\left(\frac{\partial^2 p}{\partial T^2}\right)^{-6}\frac{\partial^6 p}{\partial T^6}\Bigg]\,.\label{eq:c6}
\end{align}
The relevant numerical results are shown in Fig.~\ref{fig:c5-6}.

\section{Ratios of temperature fluctuation cumulants}
\label{app:cn-ratios}

%
\begin{figure*}[t]
\includegraphics[width=0.8\textwidth]{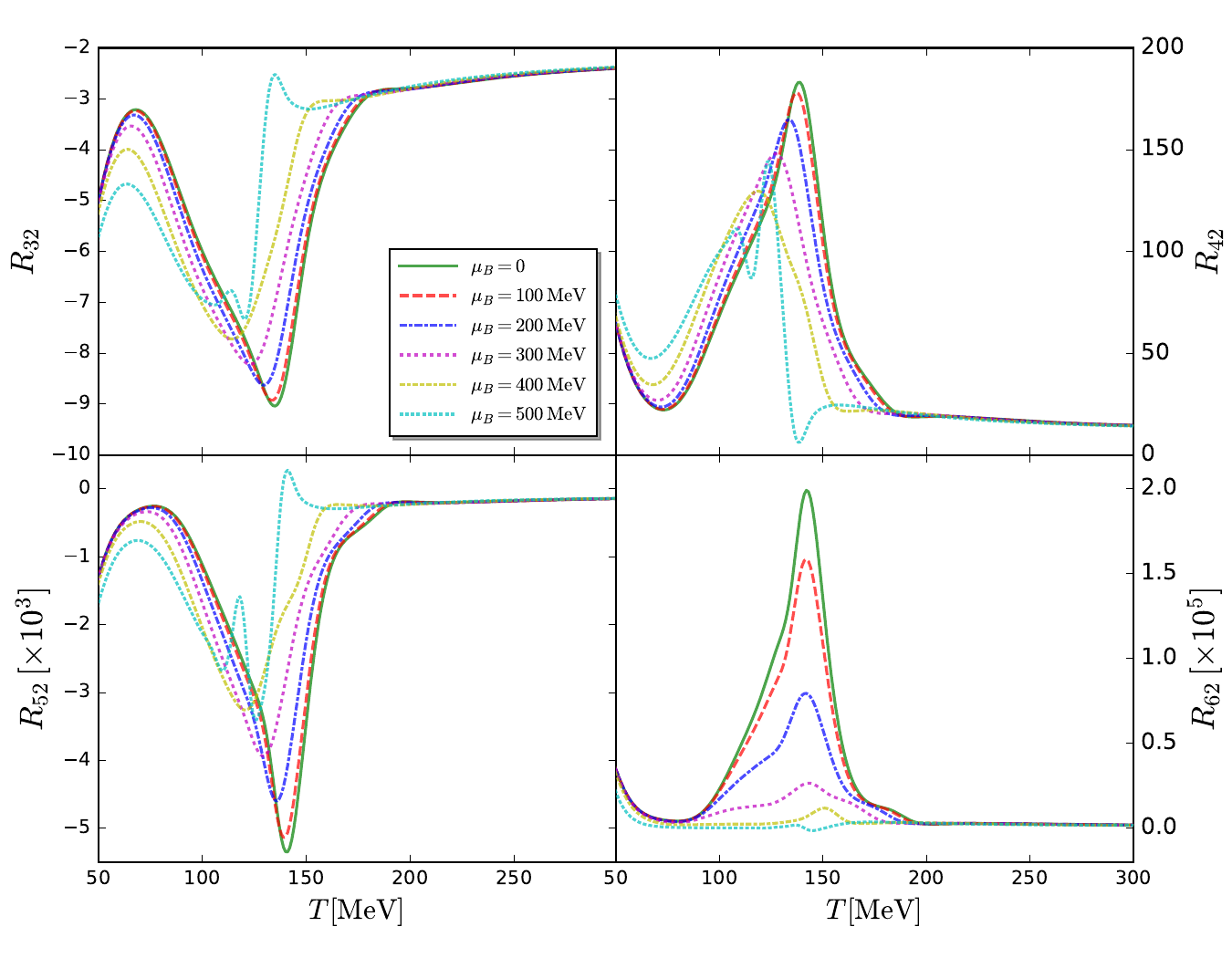}
\caption{Ratios between high-order temperature fluctuations and the variance, $R_{32}=c_3/c_2^2$, $R_{42}=c_4/c_2^3$, $R_{52}=c_5/c_2^4$, $R_{62}=c_6/c_2^5$, as functions of the temperature with several different values of baryon chemical potential.}
\label{fig:Rmn}
\end{figure*}
%

In relativistic heavy-ion collisions, the event-averaged mean transverse momentum $\langle p_{T} \rangle$ of charged particles exhibits an approximate linear dependence on the system temperature, $\langle p_{T} \rangle = a\,T$. The parameter $a$ represents the proportionality coefficient. In order to eliminate the influence from this coefficient that is not determined quite well, we instead analyze dimensionless ratios of temperature fluctuation cumulants: 
\begin{align}
    R_{32}=\frac{c_3}{c_2^2}\,, \quad R_{42}=\frac{c_4}{c_2^3}\,,\quad R_{52}=\frac{c_5}{c_2^4}\,,\quad R_{62}=\frac{c_6}{c_2^5}\,.\label{eq:R32R42}
\end{align}
where the powers of the variance in the denominators are chosen to balance the powers of $T$. The relevant ratios, presented in Fig.~\ref{fig:Rmn}, reveal that the cumulant ratios develop an increasingly rich nonmonotonic structure while exhibiting systematically reduced amplitudes with increasing $\mu_B$, reflecting competing effects where enhanced critical fluctuations near the sharpened phase boundary emerge concurrently with the overall suppression of the magnitude of temperature fluctuation. 

\section{Relation between the temperature $T$ and the mean transverse momentum $\langle p_{T} \rangle$}
\label{app:T-pT-relation}

Considering a free relativistic gas in thermal equilibrium at temperature $T$ with Boltzmann statistics, where the mass of particle, e.g., the pion $m_\pi\sim 150 $ MeV, is negligible in comparison to the typical value of the transverse momentum $p_T \sim 1$ GeV, the particle number reads
\begin{align}
    N=(2s+1)V\int \frac{d^3 p}{(2\pi)^3}e^{-p/T}\,, \label{eq:N}
\end{align}
where $s$ denotes the spin of the particle and $V$ the volume of system. One readily finds for the mean transverse momentum of particles
\begin{align}
    \langle p_{T} \rangle=\frac{1}{N}(2s+1)V\int \frac{d^3 p}{(2\pi)^3}\,p_{T}\,e^{-p/T}\,, \label{eq:pt}
\end{align}
where the momentum and the transverse momentum is related through
\begin{align}
    p=\sqrt{p_{T}^2+p_z^2}\,, \label{}
\end{align}
with the longitudinal momentum $p_z$. Combining Eqs.~\labelcref{eq:N} and \labelcref{eq:pt} we have
\begin{align}
    \langle p_{T} \rangle=\frac{\int \frac{d^3 p}{(2\pi)^3}\,p_{T}\,e^{-p/T}}{\int \frac{d^3 p}{(2\pi)^3}\,e^{-p/T}}\,. \label{eq:pt2}
\end{align}
From the dimensional analysis in Eqs.~\labelcref{eq:pt2}, one immediately arrives at
\begin{align}
    \langle p_{T} \rangle=a\, T\,, \label{eq:pt-T}
\end{align}
that is, $\langle p_{T} \rangle$ is linearly proportional to the temperature, with the proportionality coefficient $a$. Note that this is different from the non-relativistic case, where it is easy to find that $\langle p_{T} \rangle \sim T^{1/2}$. In order to determine the coefficient $a$ in Eq.~\labelcref{eq:pt-T}, one has to compute the expression in Eq.~\labelcref{eq:pt2}. The integral in the numerator in Eq.~\labelcref{eq:pt2} cannot be computed analytically, but one can expand the momentum in the exponent in powers of $p_z/p_{T}$, i.e.,
\begin{align}
    p&=p_{T}\Big(1+\frac{p_z^2}{p_{T}^2}\Big)^{1/2} \nonumber\\[2ex]
    &=p_{T}+\frac{1}{2}\frac{p_z^2}{p_{T}}+\cdots\,.\label{eq:pexpan}
\end{align}
Expanding Eq.~\labelcref{eq:pexpan} to the second order, and substituting into Eq.~\labelcref{eq:pt2}, one finds
\begin{align}
    \langle p_{T} \rangle \approx \frac{15}{16\sqrt{2}}\pi T \approx 2.08 \,T\,. \label{}
\end{align}
Note that this value is close to the exact value
\begin{align}
    a=2.356\,, \label{}
\end{align}
obtained by directly computing Eq.~\labelcref{eq:pt2} numerically.

\vfill 

\bibliography{ref-lib}%

\end{document}